%% file: Rh115pressureNMR_PRB.tex
\documentclass[aps,twocolumn,superscriptaddress,showpacs,amssymb,prb,floatfix,preprintnumbers]{revtex4-1}
\usepackage{graphicx}
\usepackage{epstopdf}
\usepackage{hyperref}

\hypersetup{colorlinks=true,linkcolor=blue,citecolor=blue,urlcolor=blue}

\newcommand{\slrr}{$T_1^{-1}$}

\begin{document}

\title{Evolution of hyperfine parameters across a quantum critical point in  CeRhIn$_5$}

\author{C. H. Lin}
\email[]{chllin@ucdavis.edu}
\author{K. R. Shirer}
\author{J. Crocker}
\author{A. P. Dioguardi}
\author{M. M. Lawson}
\author{B. T. Bush}
\author{P. Klavins}
\author{N. J. Curro}
\affiliation{Department of Physics, University of California, Davis, CA 95616}

\date{\today}

\begin{abstract}
We report Nuclear Magnetic Resonance (NMR) data for both the In(1) and In(2) sites in the heavy fermion material CeRhIn$_5$ under hydrostatic pressure.  The Knight shift data reveal a  suppression of the hyperfine coupling to the In(1) site as a function of pressure, and the electric field gradient, $\nu_{\alpha\alpha}$, at the In(2) site exhibits a change of slope, $d\nu_{\alpha\alpha}/dP$, at  $P_{c1} = 1.75$ GPa.  These changes to these coupling constants reflect alterations to the electronic structure at the quantum critical point.
\end{abstract}


\maketitle

Heavy fermion metals often exhibit strong electron-electron interactions that can be tuned across a quantum phase transition between localized f-electron magnetism and itinerant heavy-mass Fermi liquid behavior. \cite{doniach,zachreview,StewartHFreview} Between these two extremes, fluctuations associated with an antiferromagnetic quantum critical point (QCP) can give rise to non-Fermi liquid behavior and  unconventional superconductivity.\cite{ColemanQCreview,SiLocalQCP,YRSnature,Gegenwart2008} CeRhIn$_5$ is a prototypical heavy fermion compound that is antiferromagnetic below $T_N = 3.8$ K at ambient pressure, and superconducting below  a maximum $T_c=2.3$ K for hydrostatic pressures above  $P_{c1} = 1.75$ GPa.\cite{CeRhIn5ParkNJP2009}  Several measurements have uncovered changes in the basic properties of this material as pressure is tuned across this QCP.   de-Haas van Alphen (dHvA) studies revealed a discontinuous change in the Fermi surface and divergence of the effective mass across $P_{c1}$, consistent with the local $4f$ Ce moments becoming itinerant above this pressure.\cite{ShishidoRh115dHvA} Transport measurements in the paramagnetic normal phase have uncovered evidence for local quantum critical fluctuations in the vicinity of this QCP giving rise to non-Fermi liquid behavior.\cite{tuson,tusonNature2008,SiKondoDestruction2014JPSJ} Recent neutron scattering experiments indicate an incommensurate spin spiral structure in the ordered phase that persists up to $P_{c1}$ driven by frustrated  magnetic exchange interactions.\cite{NSRh1152014,Curro2000a,baoCeRhIn5INS,AnnaCeRhIn5NSpressure,Rh115NSpressure2009}

\begin{figure}
\includegraphics[width=\linewidth]{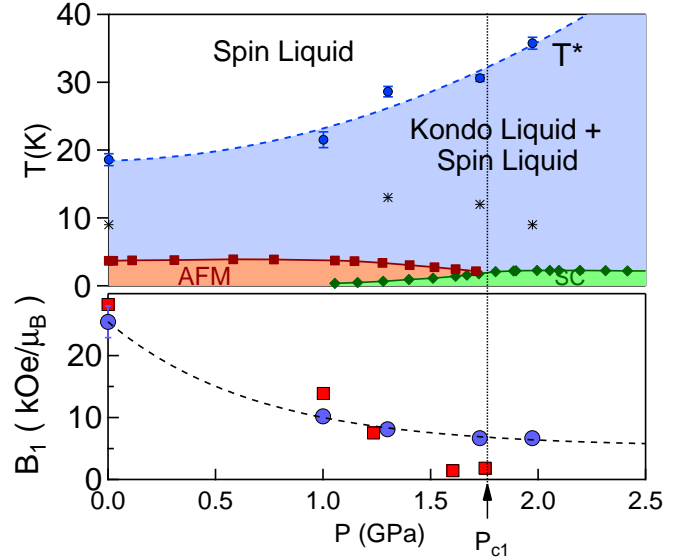}%
\caption{\label{fig:pressure_plot} (Upper panel) Pressure-temperature phase diagram of CeRhIn$_5$.  $T_N$ and $T_c$ are reproduced from \cite{CeRhIn5ParkNJP2009}. The dashed line through the $T^*$ points is a guide to the eye, and represents a crossover. $T_0$ is shown as $*$. (Lower panel) $B_1$ ($\bullet$, blue) and $10 H_{int}/\mu_{Ce}$ ($\blacksquare$, orange) vs. pressure; data from \cite{Curro2000a,kitaokaCeRhIn5pressureGapless,Rh115NSpressure2009}). The dashed line is a guide-to-the-eye, and $P_{c1}$ indicates the quantum critical point.}
\end{figure}

Nuclear magnetic resonance (NMR) has  played a central role in the study of CeRhIn$_5$ and other heavy fermion materials.\cite{KitaokaBook,Curro2009} The hyperfine interaction in these materials enable the nuclei to passively probe the static and dynamic properties of the electronic spins, and NMR is readily adapted for extreme environments such as high pressure and ultralow temperatures.  As a result, this technique is ideal for probing the microscopic response of materials across both conventional and quantum phase transitions.   Strong electron-electron interactions modify the scattering between quasiparticles near a QCP, and theoretical models can be compared with experimental measurements of NMR quantities such as the Knight shift, $K$, and the  spin-lattice relaxation rate, \slrr.\cite{SiLocalQCP,IshidaYRS2002PRL,KitaokaZhengReviewHFSC2007,Yang2009}   These studies, however, are predicated on the assumption that the hyperfine couplings do not change throughout the phase diagram.  In CeRhIn$_5$, this assumption has led to some contradictory results.  For example,  nuclear quadrupolar resonance (NQR) measurements of \slrr\ revealed a decrease in the spin fluctuations near the QCP, in contrast to transport measurements that indicate enhanced spin fluctuations.\cite{KawasakiRh115NQRpressure2001,tusonNature2008}  Further NQR experiments indicated a reduction of the ordered moment under pressure, in contrast to neutron scattering.\cite{kitaokaCeRhIn5coexistence,kitaokaCeRhIn5pressureGapless}   These discrepancies throw doubt on the validity of NMR/NQR as a viable technique to investigate quantum phase transitions.

Here we report nuclear magnetic resonance (NMR) spectral measurements of the In(1) and In(2) sites in CeRhIn$_5$ under pressure. These results reveal that the hyperfine coupling decreases with pressure, and the slope of the electric field gradient (EFG) changes at the QCP.
The anomalous behavior of the NQR internal field and \slrr\ under pressure can be fully explained by renormalizing by the pressure-dependent hyperfine coupling.  Furthermore, by analyzing the temperature dependence of the Knight shift we find that the lattice coherence temperature, $T^*$, increases with pressure, in agreement with recent predictions (see Fig. \ref{fig:pressure_plot}).\cite{YangPinesNature,YangDavidPRL,YangPinesPNAS2012,jiang14}  The EFG reflects a change in the occupation of the 5p orbitals as the Ce 4f electrons become itinerant at the QCP.

\begin{figure}
\includegraphics[width=\linewidth]{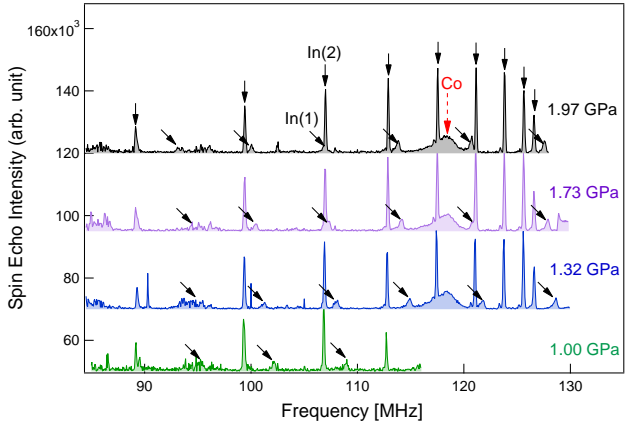}%
\caption{\label{fig:spectra} Frequency-swept NMR spectra of CeRhIn$_5$ at constant field $H_0 = 11.729$ T at various pressures.  The In(1) and In(2) resonances are indicated.  The Co resonance indicated by the arrow is a background signal and arises from the pressure cell.}
\end{figure}

High quality single crystals of CeRhIn$_5$ were synthesized using In flux  as described elsewhere.\cite{Moshopoulou200125} A crystal of mass $\sim 2$ mg was placed in a conventional piston-clamp cell  attached to a customized NMR probe, and aligned with $\mathbf{H}_0~||~\mathbf{c}$, where $H_0 = 11.729$ T.  Daphne oil was used as a pressure medium, and the pressure was calibrated using by measuring the $T_c$ of a piece of Sn located inside the pressure cell. 
$^{115}$In has spin $I=9/2$, and the multiplet is split by the large quadrupolar interaction.  At ambient pressure, the electric field gradient (EFG) at the In(1) site is $(\nu_{aa}=-3.39, \nu_{bb}= -3.39,\nu_{cc} = 6.78)$ MHz, and $(\nu_{aa}=16.665, \nu_{bb}= -12.041,\nu_{cc} = -4.625)$ MHz at the In(2) site.  At each pressure, broad frequency-swept spectra were acquired using a standard Hahn-echo pulse sequence to observe multiple quadrupolar satellites of both sites. The alignment was confirmed to within $\sim 1^{\circ}$ by observing the splitting of the quadrupolar satellites.

\begin{figure}
\includegraphics[width=\linewidth]{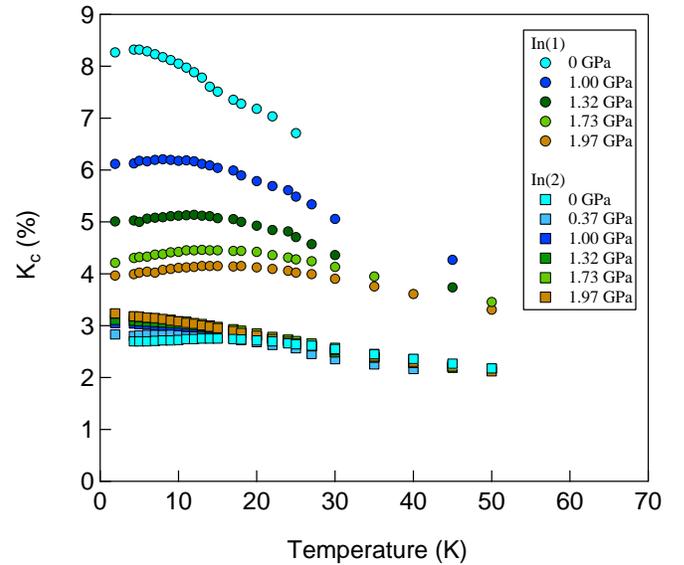}%
\caption{\label{fig:KvsT}The Knight Shift $K_c$  of the In(1)  and In(2) nuclear sites as a function of temperature and pressure for CeRhIn$\mathrm{_5}$. The point size is representative of the error bars. Data for In(1) at ambient pressure is reproduced from \onlinecite{ShirerPNAS2012}.}
\end{figure}

\begin{figure}
\includegraphics[width=\linewidth]{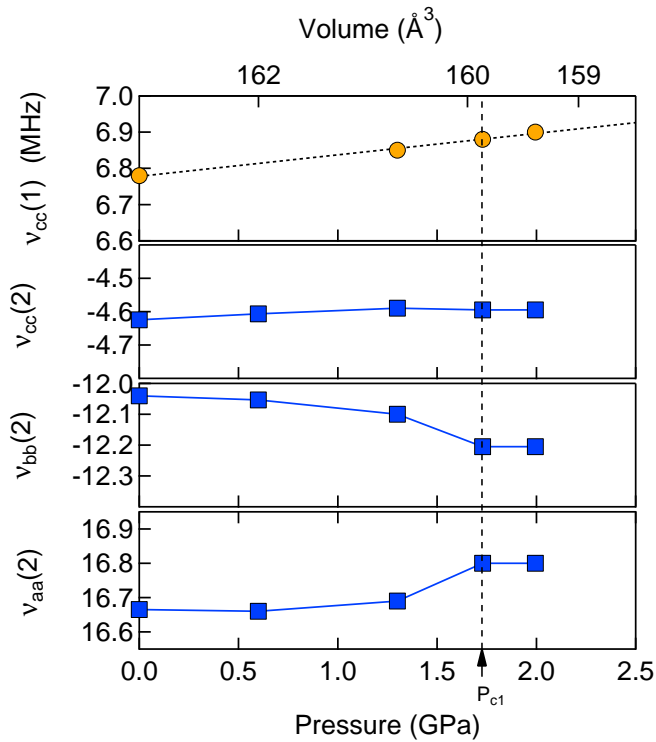}%
\caption{\label{fig:EFGvsP} EFG parameters at the In(1)  and In(2)  sites as a function of pressure for CeRhIn$\mathrm{_5}$. The point size is representative of the error bars, and the dotted line is a linear fit to the data.  Volume axis determined from elastic constants \cite{Rh115ElasticPropertiesPRB2004}.}
\end{figure}

Fig. \ref{fig:spectra} shows broad frequency-swept spectra at constant field  acquired at several different temperatures and pressures. The peak frequencies were fit by numerically diagonalizing the nuclear spin Hamiltonian to extract the Knight shifts and quadrupolar couplings. The Knight shifts, $K_c(1)$ and $K_c(2)$, of both sites are plotted in Fig. \ref{fig:KvsT}, and the EFG parameters are shown in Fig. \ref{fig:EFGvsP}. The EFG for the In(1) site shown in Fig. \ref{fig:EFGvsP} exhibits an increase with pressure. There are generally two terms that contribute to the EFG: $\nu_{\alpha\alpha} = \nu_{\alpha\alpha}^{lat} + \nu_{\alpha\alpha}^{orb}$, where the lattice contribution is given by $\nu_{\alpha\alpha}^{lat} = \beta/V$. \cite{KitaokaYbInCu41990JPSJ,Young2005,ZhengEFGcuprates1995}  Here $\beta$ is a constant, $V$ is the cell volume,  and $\nu_{\alpha\alpha}^{orb}$ is the contribution from on-site orbitals that are partially occupied.  Using the known elastic constants for this material, we convert pressure to volume and fit $\nu_{cc}(1)$ versus $V^{-1}$ to extract $\beta =840\pm40$ MHz$\cdot$\AA$^3$ and  $\nu_{cc}^{orb}(1) = 1.6\pm 3$MHz for the In(1) site.\cite{Rh115ElasticPropertiesPRB2004}  The orbital contribution arises from the 5p orbitals.\cite{IyeJPSLorbitalnematicity2015}  Unlike the In(1) EFG, the In(2) EFG does not grow smoothly with pressure but increases sharply at $P_{c1}$. The lattice constants and atomic positions evolve monotonically over this range of pressure, thus the In(2) EFG cannot be explained by a change in $\nu_{\alpha\alpha}^{lat}$.\cite{Rh115ElasticPropertiesPRB2004} We postulate that the dominant contribution to the EFG at the In(2) site arises from $\nu_{\alpha\alpha}^{orb}$; thus the discontinuous change in slope at $P_{c1}$ indicates a change in the occupations of the 5p orbitals at the In(2) site, reflecting the drastic change in the Fermi surface at the QCP.\cite{ShishidoRh115dHvA} The smaller $\nu_{cc}^{orb}$ at the In(1) may be a result of weaker hybridization at this site.\cite{HauleCeIrIn5}

\begin{figure}
\includegraphics[width=\linewidth]{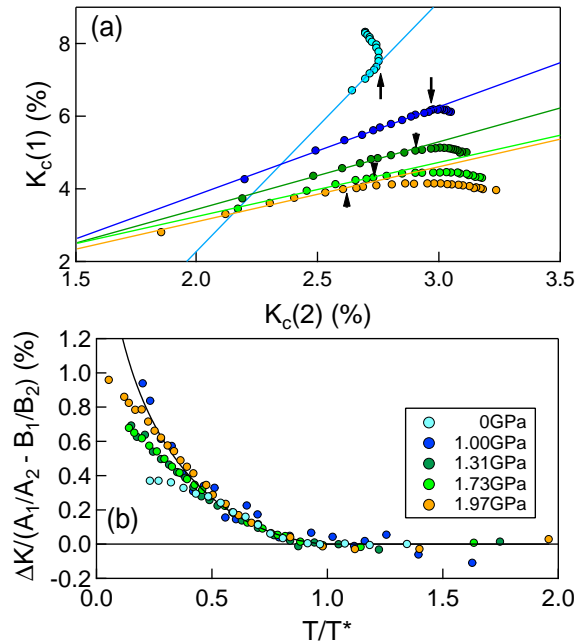}%
\caption{\label{fig:K1_vs_K2} (a) $K_c(1)$ versus $K_c(2)$ at various pressures, with temperature implicit. The solid lines are best linear fits to the high temperature regime, as discussed in the text.  The arrows indicate $T^*$, where the high temperature linear relationship breaks down. (b) $\Delta K$ versus temperature for several pressures.  The solid lines are fits to the two-fluid expression as explained in the text, and arrows indicate $T^*$.}
\end{figure}

The differences between these two sites is also evident in the Knight shift behavior.  The  shift of the In(2) site is essentially pressure-independent down to $\sim 20$ K, but below this temperature it increases slightly with increasing pressure.  The In(1) site, on the other hand, is strongly pressure dependent:  at ambient pressure it exhibits a small maximum around 8 K, but under pressure the overall scale decreases and develops a maximum that increases to about 20 K by 2 GPa.  The overall scale of $K_c(1)$ decreases by a factor of two over this range, and qualitatively begins to exhibit behavior similar to  CeCoIn$_5$, where $K_c(1)$ decreases at low temperature below a maximum.\cite{Curro2001} Fig. \ref{fig:K1_vs_K2}(a) shows $K_c(1)$ versus $K_c(2)$ with temperature implicit. For high temperatures, the shifts of the two sites are proportional to one another; but below a temperature, $T^*$, this linear relationship breaks down, as indicated by the arrows.  This anomalous behavior is a manifestation of the onset of heavy fermion coherence in these materials. In heavy fermion compounds, the hyperfine interaction at a non-f site contains two terms: $\mathcal{H}_{hyp} = \mathbf{\hat{I}} \cdot (A\cdot  \mathbf{S}_c  + {B}\cdot \mathbf{S}_f)$, where ${A}$  and ${B}$ are temperature-independent  hyperfine couplings to the conduction electron and local moment spins, $\mathbf{S}_c$ and $\mathbf{S}_f$.\cite{Curro2004}  The on-site coupling $A$ is  a Fermi-contact interaction, whereas the transferred hyperfine coupling, $B$, depends on the hybridization between the orbital with the local moment and those surrounding the nucleus in question.\cite{mila89,Meier}
In the paramagnetic state, the spins are polarized by the external field, and the Knight shift is given by ${K} = {A}{{\chi}}_{cc} + ({A}+{B})\chi_{c\!f} + {B}\chi_{f\!f}$, where $\chi_{ij} = \langle S_{i}S_{j}\rangle$ are the  components of the total spin susceptibility $\chi = \chi_{cc} + 2\chi_{cf} + \chi_{f\!f}$.  For sufficiently high temperatures, $\chi_{ff}$ is the dominant term, in which case $K \approx B\chi$.  If the strong pressure dependence of $K_c(1)$ reflected that of $\chi$, then both $K_c(1)$ and $K_c(2)$ would have similar behavior.  The only explanation for the different pressure dependence of these two quantities is that the hyperfine coupling $B_1$ to the In(1) site is suppressed with pressure.  Direct measurements of $\chi$ under pressure are unavailable, however we can compare the shift of the two sites:  $K_c(1) \approx \left(B_1/B_2\right)K_c(2)$.\cite{Shockley2011}  This linear relationship is clearly evident in Fig. \ref{fig:K1_vs_K2}(a), which reveals that the slope is a strong function of pressure.  We fit the high temperature portion to a linear function to extract the slope, assume the value $B_2 = 4.1\pm0.4$ kOe/$\mu_B$ is pressure-independent, and show the pressure dependence of $B_1$  in the bottom panel of Fig. \ref{fig:pressure_plot}.

The strong reduction in $B_1$ explains the anomalous behavior of the internal field measured by NQR, as well as the apparent suppression of spin fluctuations under pressure observed in \slrr\ measurements.  In the ordered antiferromagnetic phase, the internal field at the In(1) site, $H_{int}$,  decreases by an order of magnitude between ambient pressure and $P_{c1}$, whereas the ordered moment, $\mu_{Ce}$, only decreases by 30\% over the same range.\cite{Curro2000a,kitaokaCeRhIn5pressureGapless,AnnaCeRhIn5NSpressure,Rh115NSpressure2009}  The suppression of the internal field simply reflects the reduction of the hyperfine coupling to the In(1) site.  The lower panel of Fig. \ref{fig:pressure_plot} includes the effective coupling, $B_{eff} = 10 H_{int}/\mu_{Ce}$ as a function of pressure.  Note that the internal field is a function of the in-plane components of the tensor, whereas the Knight shift data reported here probe the out-of-plane component, which is approximately a factor of ten smaller.\cite{Curro2000a}  Nevertheless, the agreement between $B_{eff}$ and $B_1$ is compelling and suggests that all the components of the tensor are reduced with pressure.

\begin{figure}
\includegraphics[width=\linewidth]{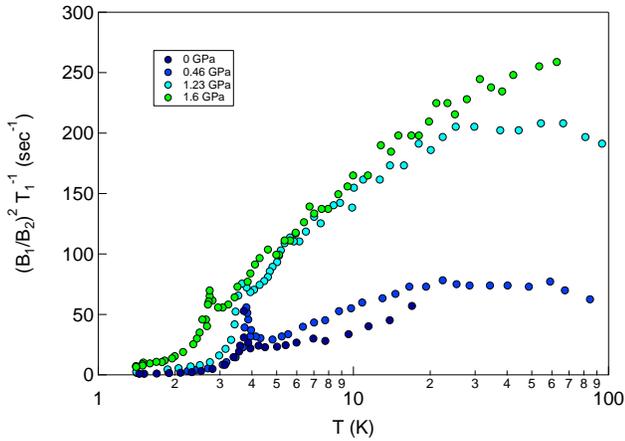}%
\caption{\label{fig:T1Tnormalized} $\left(\frac{B_{cc}^{(2)}}{B_{cc}^{(1)}}\right)^2\frac{1}{T_1}$ versus $T$ in CeRhIn$_5$ for various pressures.  The original $T_1$ data was extracted from Ref. \onlinecite{KawasakiRh115NQRpressure2001}.}
\end{figure}

The spin lattice relaxation rate also depends on the hyperfine coupling through the form factor:
\begin{equation}
\frac{1}{T_1T} = \frac{\gamma^2k_B}{2}\lim_{\omega\rightarrow 0}\sum_{\mathbf{q}}F^2(\mathbf{q})\frac{\chi''(\mathbf{q},\omega)}{\omega},
\end{equation}
where $\gamma$ is the gyromagnetic ratio and the form factor $F(\mathbf{q})$ is given by the hyperfine coupling in $\mathbf{q}$ space.\cite{Curro2000a}  The quantity $\frac{1}{B_1^2}\frac{1}{T_1T}$ therefore removes the hyperfine coupling dependence and is a direct measurement of the dynamical electron spin susceptibility, $\chi''(\mathbf{q},\omega)$. As shown in Fig. \ref{fig:T1Tnormalized}, this quantity increases with increasing pressure, reflecting the growth of critical fluctuations as the system approaches the QCP, consistent with transport measurements.\cite{tusonNature2008}

Below $T^*$, the conduction electrons and the local moments become entangled, and $\chi_{cf}$ grows in magnitude.  Since $A\neq B$ in general, the linear relationship between $K$ and $\chi$ breaks down, and the plot of $K_c(1)$ versus $K_c(2)$ (Fig. \ref{fig:K1_vs_K2}(a)) exhibits a change of slope.  This feature enables us to directly track $T^*$ as a function of pressure, as shown in Fig. \ref{fig:pressure_plot}.  Fig. \ref{fig:K1_vs_K2}(b) shows the quantity:
\begin{eqnarray}
\nonumber  \Delta K &=& \left(K_c(1) - (B_1/B_2) K_c(2)\right)/\left(A_1/A_2  - B_1/B_2\right)\\
&=& A_2 \left(\chi_{cf} +\chi_{cc}\right)
\label{eqn:deltaK}
\end{eqnarray}
versus $T/T^*$.\cite{CeIrIn5HighFieldNMR} Here we assume that the Fermi-contact  terms $A_1$ and $A_2$ do not change with pressure, and $A_1/A_2 = 2.83$.\cite{Curro2004}  The data clearly scale with $T/T^*$ down to a temperature, $T_0$, which is shown in Fig. \ref{fig:pressure_plot}. The solid line shows the two-fluid expression $K_{HF}(T) \sim (1 - T/T^*)^{3/2}[1+\ln(T^*/T)]$.\cite{YangDavidPRL} $T^*$ indicates a crossover temperature scale, below which the local f moments become entangled with the conduction electrons, giving rise to two-fluid behavior in which the material exhibits both local moment and heavy electron behavior simultaneously.  These data agrees with indirect measurements of the coherence temperature under pressure extracted from resistivity measurements.\cite{YangPinesNature} Our results also verify  a predictive standard model for heavy-fermion systems, in which $T^*$ sets the scale for collective hybridization and the emergence of a Kondo liquid.\cite{Yang2011} The increase in $T^*$ reflects an increase in the hybridization between the f-moments and the conduction electrons.  Recent quantum Monte Carlo (QMC) calculations of the periodic Anderson lattice model revealed a systematic evolution of $\chi_{cc}$, $\chi_{cf}$ and $\chi_{ff}$ as a function of the hybridization parameter, $\mathcal{V}$, representing hopping between the local moment sites and the conduction electron sites.\cite{jiang14}  Both the logarithmic behavior of $\Delta K$ and the increase in $T^*$ with $\mathcal{V}$ are well captured by these calculations. Our results therefore provide clear evidence that pressure increases the hybridization, presumably by increasing the orbital overlap as the lattice spacing decreases (the Ce-Ce distance decreases by $\sim$0.8\% by 2 GPa).\cite{Rh115ElasticPropertiesPRB2004}  $T_0$ probably signals the onset of relocalization, in which the hybridized quasiparticles partially localize prior to the onset of long range antiferromagnetism.\cite{Warren2011,YangPinesPNAS2012}

An increase in hybridization  offers a natural explanation for the strong pressure dependence of the transferred hyperfine coupling, $B_1$, to the In(1) site.  The Ce 4f hybridizes with the 5s and 5p orbitals of both the In(1) and In(2) sites,\cite{HottaPRB2003} which provides a mechanism for the transferred hyperfine interaction between the In nuclear spin and the Ce moments.   As the localized 4f state with strong spin-orbit coupling becomes more delocalized, the dipolar field it creates at the In(1) decreases. Detailed electronic structure calculations of the hyperfine coupling are unavailable, but should be able to capture these trends.

Surprisingly, the hyperfine coupling is strongly pressure dependent at the In(1) site, but not at the In(2).  This discrepancy suggests that the In(2) orbitals are already strongly hybridized at ambient pressure.  Electronic structure calculations do indicate a stronger hybridization at the In(2) site in isostructural CeIrIn$_5$.\cite{HauleCeIrIn5}  This interpretation also explains why the EFG at the In(2) is more sensitive to the change in the Fermi surface than the In(1) site.   Our results suggest that that two-dimensional Ce-In(1) plane thus plays a key role in the emergence of superconductivity in the CeMIn$_5$ series (M = Co, Rh, Ir), which only emerges once the in-plane hybridization has increased sufficiently.  In conclusion, we note that NMR/NQR studies of quantum critical behavior under pressure must take into account the possibility for pressure-dependent hyperfine couplings, which can both alter the interpretation of key results and shed light on microscopic details of the electronic structure.

We thank  D. Pines, Y-f. Yang, M. Jiang, A. Benali, and R. Scalettar for enlightening discussions.  This work was supported by the UC Office of the President and  the National Nuclear Security Administration under the Stewardship Science Academic Alliances program through DOE Research Grant \#DOE DE-FG52-09NA29464.

\input{Rh115pressureNMR_PRB.bbl}
\end{document}

%% file: Rh115pressureNMR_PRB.bbl
%